\documentclass[11pt]{article}

\usepackage{amsmath,amssymb,amsfonts,color,graphics,graphicx,amscd,amsfonts,epsf,epsfig}

\usepackage{color}

\newcommand{\eq} {equation}
\newcommand{\eqa} {eqnarray}
\newcommand{\NN} {\mbox {$\nonumber$}}

\newcommand{\ba}{\begin{eqnarray}}
\newcommand{\ea}{\end{eqnarray}}

\newcommand{\nt}{\nonumber\\}

\allowdisplaybreaks[1]

\date{}
\begin{document}

\begin{flushright} 
YITP-13-50

KEK-TH-1638

\end{flushright} 

\vspace{0.1cm}

\begin{center}
  {\LARGE\bf   A new look at instantons and large-$N$ limit}

\end{center}

\vspace{0.2cm}

\begin{center}

{\large
Tatsuo A{\sc zeyanagi},${}^{a\,}$\footnote{e-mail: 
azey@physics.harvard.edu}
Masanori H{\sc anada},${}^{bcd\,}$\footnote{e-mail: 
hanada@yukawa.kyoto-u.ac.jp}
Masazumi H{\sc onda},${}^{bd\,}$\footnote{e-mail: 
mhonda@post.kek.jp}\\[+3pt]
Yoshinori M{\sc atsuo}${\,}^{d\,}$\footnote{e-mail: 
ymatsuo@post.kek.jp}
and 
Shotaro S{\sc hiba}${}^{d\,}$\footnote{e-mail: 
sshiba@post.kek.jp}  
}
\vspace{0.6cm}

${}^a$
{\it Center for the Fundamental Laws of Nature, Harvard University,\\
 Cambridge, Massachusetts 02138, USA}\\

${}^b$
{\it Yukawa Institute for Theoretical Physics, Kyoto University,\\
Kitashirakawa Oiwakecho, Sakyo-ku, Kyoto 606-8502, Japan} \\

${}^c$
{\it The Hakubi Center for Advanced Research, Kyoto University, \\
  Yoshida-Ushinomiya-cho, Sakyo-ku, Kyoto 606-8501, Japan}

${}^d$ {\it KEK Theory Center, High Energy Accelerator Research Organization (KEK), \\
Oho, Tsukuba, Ibaraki 305-0801, Japan}\\

\end{center}

\setcounter{footnote}{0}

\vspace{1.2cm}

\begin{center}
  {\bf Abstract}
\end{center} 
We analyze instantons in the very strongly coupled large-$N$ limit ($N\to\infty$ with $g^2$ fixed) 
of large-$N$ gauge theories, 
where the effect of the instantons remains finite. 
By using the exact partition function of four-dimensional ${\cal N}=2^*$ gauge theories as a concrete example,
we demonstrate that each instanton sector
in the very strongly coupled large-$N$ limit  is related to 
the one in the 't Hooft limit ($N\to\infty$ with $g^2N$ fixed) through a simple analytic continuation.
Furthermore we show the equivalence between the instanton partition functions of a pair of  large-$N$ gauge theories 
related by an orbifold projection. 
This can open 
up a new way to analyze the partition functions of low/non-supersymmetric theories.
We also discuss implication of our result to gauge/gravity dualities for M-theory 
as well as a possible application to large-$N$ QCD.

\newpage

\section{Introduction}

The 't Hooft limit of large-$N$ gauge theories \cite{'tHooft:1973jz}, $N\to\infty$ with the 't Hooft coupling $\lambda=g^2N$ fixed,  
has been playing a prominent role in various fields of theoretical physics. 
Around the 't~Hooft limit, there exists a $1/N$ expansion which rearranges the Feynman diagrams in a geometric manner. 
In the 't Hooft limit, only the planar diagrams survive and drastic simplification takes place.  
When it comes to instantons, however, it is not clear whether the 't Hooft limit is an appropriate playground, 
because the instanton action grows as $1/g^2=N/\lambda \sim N$, providing an exponential suppression factor of the form $e^{-N}$.\footnote{
By taking quantum effect into account, the weight can behave as $e^{-f(\lambda)/g^2}$, where $f(\lambda)$ satisfies $f(\lambda)>0$ at weak coupling. 
Then if $f(\lambda)$ becomes zero, the instantons can give a nontrivial contribution. See a nice review \cite{Marino:2012zq} for details. 
} 

In this letter, we demonstrate that a certain information of the instantons can be extracted from the 't Hooft limit. 
The starting point is to consider a more general large-$N$ limit, $N\to\infty$ with $\lambda\sim N^p$ $(p>0)$;  
here we call it {\it the very strongly coupled large-$N$ limit}. 
As a special case, it contains a large-$N$ limit with fixed $g^2$ ($p=1$), where the instanton action is of order one 
and hence the instanton effect is not suppressed.  
In this letter, we also call this special case simply by the very strongly coupled limit,  
unless otherwise stated.
In \cite{Azeyanagi:2012xj,Fujita:2012cf}, it has been shown that, in the zero instanton sector, 
various properties in the 't Hooft limit are inherited to the the very strongly coupled large-$N$ limit.
The argument is very simple for theories with gravity duals. As the simplest example, let us consider the four-dimensional ${\cal N}=4$
$SU(N)$ supersymmetric Yang-Mills theory. 
Through the AdS/CFT correspondence \cite{Maldacena:1997re}, this theory is dual to 
classical type IIB supergravity on $AdS_5\times S^5$ background provided that $N\to\infty$ with 
$1\ll \lambda\ll N$
is satisfied. Here the first and the second inequalities respectively guarantee the string length and loop corrections are negligible.
Usually one takes the 't Hooft limit first and then sends the 't Hooft coupling to large but still of order one (i.e. $\lambda$ does not scale with $N$).  
It is however not really needed for the AdS/CFT to be valid; the very strongly coupled large-$N$ limit with $0<p\le1$ is also described by the classical supergravity. 
In other words, the correct results in the very strongly coupled large-$N$ limit can be obtained by the analytic continuation 
from the 't Hooft limit, at least when $p\le 1$. 
The same property holds in other theories too, 
even without gravity duals or sometimes even at $p>1$ \cite{Azeyanagi:2012xj}.

In this letter we generalize the argument of  \cite{Azeyanagi:2012xj} to take into account the instanton effect. 
As we have mentioned, the instanton effect is in general exponentially suppressed in the 't Hooft limit, 
while it is of $O(1)$ when $g^2$ is fixed. 
If we consider a fixed instanton sector, however, calculations in the 't Hooft limit still make sense, 
because the 't Hooft expansion is allowed in this sector as in the zero-instanton sector.     
Below we argue that the property of the instantons in the very strongly coupled large-$N$ limit
can be extracted from such calculations.

For this purpose,  we consider the free energy of four-dimensional
${\cal N}=2^*$ $U(N)$ gauge theory as a concrete example.  
The instanton partition function is given by Nekrasov's formula at any $g$ and $N$, and we can confirm the validity of our conjecture 
by taking appropriate limits.  
It strongly suggests that various nice properties in the 't Hooft limit are smoothly extended to the very strongly coupled large-$N$ limit, even in the sectors with non-zero instanton numbers.
As an example, we show that the large-$N$ orbifold equivalence \cite{Kachru:1998ys,Bershadsky:1998cb,Kovtun:2004bz} holds in each instanton sector.  
More concretely, we consider ${\cal N}=2^*$ $U(kN)$ gauge theory and 
${\cal N}=2$ $[U(N)]^k$ necklace quiver gauge theory related by an orbifold projection, 
and show the matching of the contribution to the free energies from the instanton sectors. 
In a similar manner, we can also consider more general orbifold projections mapping ${\cal N}=2$ theories to ${\cal N}<2$ theories. 
This can allow us to analyze the instanton effects in low/non-supersymmetric theories.

This paper is organized as follows. 
In Sec.~\ref{sec:from_weak_to_strong} we introduce ${\cal N}=2^*$ gauge theories and clarify
the connection between the 't Hooft limit and the very strongly coupled large-$N$ limit.  
In Sec.~\ref{sec:orbifold_equivalence} we explain the orbifold equivalence, show how it is generalized 
to the instanton sectors, and then confirm the validity for a specific example.
Sec.~\ref{sec:discussion} is devoted for discussions on our results and future directions.

\section{From $g^2N$ fixed to $g^2$ fixed}\label{sec:from_weak_to_strong}
As a concrete setup,
let us consider the free energy of four-dimensional ${\cal N}=2^*$ $U(kN)$ gauge theory on $S^4$ with a unit radius 
($k$ and $N$ are integers). 
This theory is realized as a deformation of ${\cal N}=4$ $U(kN)$ supersymmetric Yang-Mills theory by adding a mass term to 
the ${\cal N}=2$ hypermultiplet part. We denote the mass parameter by $m$ (see e.g.~\cite{Pestun:2007rz})
and fix it to be of order one. 

The partition function of ${\cal N}=2^*$ $U(kN)$ gauge theory with the gauge coupling $g_p$ is given by 
the following integral expression with respect to the ``eigenvalues" (or equivalently the Coulomb parameters)
$a_i$ $(i=1,2,\cdots, kN)$~\cite{Pestun:2007rz,Nekrasov:2002qd}:
\begin{\eq}
\mathcal{Z}_{\mathcal{N}=2^\ast}
=\int d^{kN} a \Bigl(\prod_{\substack{i,j=1 \\[+1pt] i<j}}^{kN} (a_i -a_j )^2 \Bigr) Z^{\rm (pert)}_{\mathcal{N}=2^\ast}(a_i, m)
  \bigl|Z^{\rm (inst)}_{\mathcal{N}=2^\ast}(a_i, \tilde{m})\bigr|^2 \exp\left(-\frac{8\pi^2}{ g_p^2} \displaystyle{\sum_{i=1}^{kN}} a_i^2\right) . 
\end{\eq}
Here the perturbative one-loop contribution $Z^{\rm (pert)}_{\mathcal{N}=2^\ast}(a_i, m)$ and the instanton contribution 
$Z^{\rm (inst)}_{\mathcal{N}=2^\ast}(a_i, \tilde{m})$
are given by 
\begin{\eqa} 
&& Z^{\rm (pert)}_{\mathcal{N}=2^\ast}(a_i, m) =\prod_{\substack{i,j=1 \\[+1pt] i\neq j}}^{kN} Z_{\rm vec}^{(\rm pert)}(a_i -a_j ) 
 \prod_{\substack{i,j=1 \\[+1pt] i\neq j}}^{kN} Z_{\rm mat}^{(\rm pert)} (a_i -a_j, m)\,, \nt
&& Z_{\rm vec}^{(\rm pert)}(a_i -a_j ) =  H(i(a_i -a_j )) \,, \nt
&& Z_{\rm mat}^{(\rm pert)}(a_i -a_j, m) 
=  e^{(1+\gamma)m^2} \left[H(i (a_i -a_j +m)) H(i (a_i -a_j -m))\right]^{-\frac{1}{2}} \,,
\end{\eqa}
and 
\begin{\eqa}
&& Z^{\rm (inst)}_{\mathcal{N}=2^\ast}(a_i, \tilde{m})
= \sum_{Y=\{Y_1,\cdots, Y_{kN}\}}  
e^{-\frac{8\pi^2|Y|}{g_p^2}}
 \prod_{i,j=1}^{kN} Z_{\rm vec}^{\rm (inst)} (a_i -a_j ;Y_i ,Y_j ) 
                                 Z_{\rm mat}^{\rm (inst)} (a_i -a_j ;Y_i ,Y_j ;\tilde{m} )\,, \NN\\
&& Z_{\rm vec}^{\rm (inst)} (a_i -a_j ;Y_i ,Y_j )
=  \prod_{s\in Y_i} [E(a_i -a_j ;Y_i ,Y_j ,s)]^{-1} \prod_{t\in Y_j} [2 -E(a_j-a_i ;Y_j ,Y_i ,t))]^{-1}\,, 
\label{eq:inst_part} \\
&& Z_{\rm mat}^{\rm (inst)} (a_i-a_j ;Y_i ,Y_j ;\tilde{m} )
=  \prod_{s\in Y_i}   (E(a_i-a_j ;Y_i ,Y_j ,s) -\tilde{m}) 
   \prod_{t\in Y_j}(2 -E(a_j-a_i ;Y_j,Y_i ,t) -\tilde{m}) \,, \nonumber
\end{\eqa}
with 
\begin{\eqa}
&&H(z)
= e^{-(1+\gamma )z^2} \prod_{n=1}^\infty \left( 1 -\frac{z^2}{n^2} \right)^n e^{\frac{z^2}{n}} \,,\quad
 \tilde{m} = im+1 \,, \NN\\ 
&& E(a_i -a_j ,Y_i ,Y_j ,s) = -h_{Y_j}(s)  +(v_{Y_i}(s) +1) +i(a_j -a_i )  \,. 
\end{\eqa} 
Here $\gamma$ is Euler's constant. 
Each instanton configuration is labeled
by a set of Young tableaux 
$Y=(Y_1,\cdots,Y_{kN})$, where the total number of boxes $|Y|=\sum_{i=1}^{kN}|Y_i|$ in the tableaux corresponds to the instanton number
($Y_i$ can simply be empty, $\emptyset$). 
The contributions from the instantons and anti-instantons to the partition function are given by
 $Z^{\rm (inst)}_{\mathcal{N}=2^\ast}$ and its complex conjugate, respectively.  
The parameter $s =(s_h ,s_v )$ labels the position of a box ($s_h$-th column and $s_v$-th row) in a given Young tableau $Y_i$.
For a given $s$, $h_{Y_i} (s)$ and $v_{Y_i} (s)$ are defined by
$h_{Y_i} (s) = \nu_{s_v}^\prime (Y_i) -s_h$ and $v_{Y_i} (s) =\nu_{s_h} (Y_i) -s_v$, 
where $\nu_{s_v}^\prime (Y_i)$ and $\nu_{s_h} (Y_i)$ are length of $s_v$-th row and $s_h$-th column in the 
Young tableau $Y_i$, respectively.

To show the validity of the analytic continuation from the 't Hooft limit to the very 
strongly coupled large-$N$ limit, we first consider the zero-instanton sector.  
It turns out that analysis in the 't Hooft limit  \cite{Buchel:2013id} is straightforwardly generalized to the very strongly coupled large-$N$ limit,   
because the saddle point method used in \cite{Buchel:2013id} is valid  
both in the 't Hooft limit and in the very strongly coupling limit,
as long as $g_{p}^2\ll 1$ or equivalently $\lambda_p=g_p^2(kN)\ll N$ (that is, $g^2_p$ can be of $O(N^0)$). 
If we further assume $\lambda_p \gg 4\pi^2 m^2 /(m^2 +1)$, 
the spectral density $\rho(a)=\lim_{N\to\infty}\sum_{i=1}^{kN}\delta(a-a_i)/(kN)$ obeys 
the semi-circle law given by  
\begin{\eq}
\rho (x) = \frac{2}{\pi\mu^2} \sqrt{\mu^2 -x^2}\,,
\end{\eq}
where $\mu =\sqrt{\lambda_p (m^2+1)}/(2\pi)$\,.
Then the free energy at the leading order of the large-$N$ limit is given by 
\begin{eqnarray}
F^{\rm (pert)}
= -(kN)^2(1+m^2)\left(\frac12\log\frac{\lambda_p(1+m^2)}{16\pi^2}+\frac{1}{4}+\gamma
\right).
\end{eqnarray} 
We can thus confirm that the free energy in the zero-instanton sector takes 
the same expression in the 't Hooft limit and the very strongly coupled large-$N$ limit.\footnote{
In this case, there is no singularity separating two limits. For the case with singularities, see 
\cite{Azeyanagi:2012xj}. }

As a next step, we move to the instanton part.\footnote{
Analysis in the limit $m\rightarrow\infty ,\lambda\rightarrow 0$ has been done in \cite{Russo:2012ay}.
}
Here we consider a fixed instanton sector, 
in which the number of the instantons and anti-instantons is finite.  
Let us denote eigenvalues with which non-empty Young tableaux are associated (i.e. eigenvalues describing instantons) by $b_i$, 
while the other eigenvalues, for which the corresponding Young tableaux are empty, by $a$. 
The instanton/anti-instanton contribution to the free energy is $O(g_p^{-2})$, which is sub-leading in 
the $1/N$ expansion.   
Then the spectral density of $a$ is the same as the one in the zero-instanton sector at the leading order of the large-$N$ limit. 
Furthermore, in the instanton part, the interaction between $b_i$'s is negligible compared to the one between $b_i$ and $a$. 
The contribution to the free energy  
from the instanton sector labeled by the Young tableaux $Y$,  
$F^{\rm(inst)}_Y=-\log (\mathcal{Z}_{Y}^{\rm}/\mathcal{Z}_{\emptyset})$ 
($\mathcal{Z}_{Y}$: the partition function on the instanton background labeled by $Y$,  
  $\mathcal{Z}_{\emptyset}$: the partition function on the zero-instanton background),
is therefore given by 
\begin{align}
 F_Y^{\rm (inst)}
 =
 -kN  \sum_{b_i \in \text{(inst)}} 
 \int da\,\rho(a) 
  \log Z_Y^\text{(inst)}(b_i,a,Y_i,\emptyset)
  +
    \Delta_{\rm(pert)}(b_i) 
 \label{eq:instanton_free_energy_U(kN)}
\end{align}
at the leading order at large-$N$.
Here $Z_Y^{\rm (inst)}$ stands for the contribution of the instanton configuration labeled by $Y$ on the right hand side of the first line of \eqref{eq:inst_part}, and 
$\Delta_{\rm (pert)}(b_i)$ is the change of the perturbative part, 
which is zero as long as $-\mu\le b \le\mu$ and is positive otherwise.  
In the same manner, the contribution from anti-instantons, labeled
by Young tableaux $Y'$, 
is given by 
\begin{align}
 F_{Y'}^{\rm (anti-inst)}
 =
 -kN  \sum_{b'_i \in \text{(inst)}} 
 \int da\,\rho(a)
  \left(\log Z_{Y'}^\text{(inst)}(b'_i,a,Y'_i,\emptyset)\right)^\ast
  +
  \Delta_{\rm(pert)}(b'_i)\,, 
 \label{eq:anti-instanton_free_energy_U(kN)}
\end{align}
where $b'_i$ are eigenvalues describing anti-instantons. 
Strictly speaking, the total free energy does not decompose to a sum of \eqref{eq:instanton_free_energy_U(kN)} and \eqref{eq:anti-instanton_free_energy_U(kN)} 
when both $Y_i$ and $Y'_i$ are not empty for some $i$.
In the present case, however, contributions from such configurations are sub-leading. 

By substituting the semi-circle law for the instanton contribution, one therefore obtains 
the same expression for the contribution to the free energy from a fixed (finite-)instanton sector
in the 't Hooft and the very strongly coupled large-$N$ limit.  
For example, one-instanton contribution to the free energy labeled 
by $Y=(\square,\emptyset,\cdots,\emptyset)$, is 
\begin{eqnarray}
F_Y^{\rm(inst)} \!&\simeq&\! 
\frac{8\pi^2}{g_p^2} 
-\log\left[\frac{(1-\tilde{m})^2}{(2-\tilde{m})\tilde{m}}\right]
\nonumber\\
& &
-kN\int da\, \rho(a) \log\left[\frac{(2-i(b-a)-\tilde{m})(i(b-a)-\tilde{m})}{(2-i(b-a))i(b-a)}\right]
 +\Delta_{\rm(pert)}(b). \label{eq:fy}
\end{eqnarray} 
If one further wants to integrate with respect to $b$, the calculation becomes 
simple when $g^2_p\ll 1$; in this case, $-\mu < b <\mu$, where $\Delta_{\rm (pert)}(b)$ vanishes, 
give dominant contribution. The result of the integration is 
\ba
&&\log {\rm Re}\int_{-\mu}^\mu  db\,\exp\left(
kN\int da\, \rho(a) \log\left[\frac{(2-i(b-a)-\tilde{m})(i(b-a)-\tilde{m})}{(2-i(b-a))i(b-a)}\right]
\right) = O(\log g_p^2)
\, ,
\ea 
and hence only the first two terms on the right hand side of  \eqref{eq:fy} survive.
We note that, from the dual gravity point of view, it is natural to expect that the contribution from the higher genus in the perturbative part is less important at least when $g_p^2\ll1$.
We also notice that, although genus one diagrams in the perturbative sector may give comparable contribution to the free energy as the one from the instantons when $g^2_p\sim 1$, they are common to all the sectors with 
finite instanton numbers at the leading order of the large-$N$ limit and hence the comparison of the instanton actions still makes sense.

It is straightforward to take into account multi-instanton configurations to the free energy and sum them up at the leading order of the large-$N$ limit.
To understand it let us remind that the free energy for generic tableaux $Y$ 
decomposes to a sum of contribution from each eigenvalue as 
$F^{\rm (inst)}_Y \simeq \sum_{Y_i\neq\emptyset}F^{\rm (inst)}_{Y_i}$,
because the interaction between the instantons, $b_i$'s, is negligible. 
The free energy therefore becomes 
\begin{eqnarray}
F
=
- \log \mathcal{Z}_{\emptyset}
+kN\log\left|1+\sum_{\tilde{Y}}e^{-F^{\rm (inst)}_{\tilde{Y}}}\right|^2,  
\label{total_free_energy}
\end{eqnarray}  
where $\tilde{Y}$'s stand for a subset of $Y$'s with $Y_1\neq\emptyset$, $Y_2=Y_3=\cdots=Y_N=\emptyset$. 
\section{Orbifold equivalence in the instanton sector}\label{sec:orbifold_equivalence}
In this part we consider the orbifold equivalence
which relates the `parent theory' to its `orbifold daughter theory' obtained by an orbifold projection.  
The statement of the usual orbifold equivalence for the free energy is as follows; in the 't Hooft limit, 
by setting the 't Hooft couplings of the parent and daughter theories, $\lambda_p=g_p^2(kN)$ and $\lambda_d=g_d^2N$ respectively,  to be the same, 
$\lambda_p=\lambda_d\equiv\lambda$, the free energies of these two theories are related by $F_p(\lambda,N) = kF_d(\lambda,N)$. 
In \cite{Azeyanagi:2012xj,Fujita:2012cf} it was generalized to the very strongly coupled large-$N$ limit. 
Around the zero-instanton vacuum, this equivalence can be proven by matching the planar diagrams in the two theories. 
It is natural to expect that the same argument holds around the vacuum with a non-zero instanton number in the 't Hooft limit, 
and it can be extended to the very strongly coupled large-$N$ limit. 
Below we demonstrate that this equivalence does hold in the instanton sectors.
Our argument below 
can apply
both to the 't Hooft limit and to the very strongly coupled large-$N$ limit.

As a concrete example, we take four-dimensional ${\cal N}=2^*$ $U(kN)$ gauge theory as a parent theory, 
and relate it to the ${\cal N}=2$ $[U(N)]^k$ necklace quiver gauge theory
by an appropriate orbifold projection. 
Here we consider an orbifold projection preserving 4d ${\cal N}=2$ supersymmetry, so that the free energy of the daughter theory can also be calculated analytically. 
The projection condition for the fields in the ${\cal N}=2$ vector multiplet, such as a gauge field $A_\mu$, is given by   
$\Omega A_\mu \Omega^{-1} =A_\mu$, while the fields in the ${\cal N}=2$ hypermultiplets (we denote them symbolically as $\Phi$) are projected as  
$\Omega \Phi \Omega^{-1} = \omega^{-1} \Phi$, 
where 
$\Omega = {\rm diag}(\omega \otimes \mathbf{1}_{N\times N} ,\omega^2 \otimes \mathbf{1}_{N\times N},\cdots , \omega^k \otimes\mathbf{1}_{N\times N})$ 
and
$\omega =\exp(2\pi i/k)$.

We denote the gauge fields of the parent and daughter theories by ${\cal A}_\mu$ and $A_\mu^{(\alpha)}$, respectively, 
where $(\alpha)$ is the label of the $k$ $U(N)$ gauge groups ($\alpha=1,2,\cdots,k$).  
Let us denote the instanton solutions of the $U(kN)$ and $[U(N)]^k$ by $\bar{{\cal A}}_\mu$ and $(\bar{A}_\mu^{(1)},\cdots,\bar{A}_\mu^{(k)})$.  
Then, in the $U(kN)$ theory, as a special case we have a block-diagonal configuration $\bar{{\cal A}}_\mu={\rm diag}(\bar{A}_\mu^{(1)},\cdots,\bar{A}_\mu^{(k)})$, 
that is, the instanton moduli of the $[U(N)]^k$ theory is a subset of that of the $U(kN)$ theory. 
Note that this configuration is projected to $(\bar{A}_\mu^{(1)},\cdots,\bar{A}_\mu^{(k)})$ by the orbifold projection. 
If the instanton number of the configuration $\bar{A}_\mu^{(\alpha)}$ is $l_\alpha$, the instanton number is $\sum_{\alpha=1}^k l_\alpha$ in both theories
(From here on we consider only instantons for notational simplicity, but anti-instantons can be incorporated straightforwardly).

The correspondence between the classical actions of the parent and daughter theories, 
$S_p^{\rm cl}$ and $S_d^{\rm cl} $ respectively, is easy to see: they are calculated as   
\begin{eqnarray}
&&S_p^{\rm cl}
=
\frac{8\pi^2}{g_p^2}\sum_{\alpha=1}^k l_\alpha
=
\frac{8\pi^2 kN}{\lambda}\sum_{\alpha=1}^k l_\alpha, \quad S_d^{\rm cl}
=
\frac{8\pi^2}{g_d^2}\sum_{\alpha=1}^k l_\alpha
=
\frac{8\pi^2 N}{\lambda}\sum_{\alpha=1}^k l_\alpha,  
\end{eqnarray}
respectively and thus $S_p^{\rm cl} = k S_d^{\rm cl}$ is satisfied.  

As a next step, we show the agreement of quantum corrections.
The partition function of the daughter theory is \cite{Pestun:2007rz,Nekrasov:2002qd}
\begin{\eqa}
\mathcal{Z}_{[U(N)]^k}
&=& \int 
 \Bigl( \prod_{\alpha =1}^k  d^N a^{(\alpha)}  \prod_{\substack{i,j=1 \\[+1pt] i<j}}^N (a_i^{(\alpha )} -a_j^{(\alpha )} )^2  \Bigr)
Z_{[U(N)]^k}^{\rm (pert)} \Bigl| Z_{[U(N)]^k}^{\rm (inst)}\Bigr|^2 \exp\left(-\frac{8\pi^2}{g_d^2} \sum_{\alpha=1}^k \sum_{i=1}^N (a_i^{(\alpha )})^2\right) ,
\nonumber \\
\end{\eqa}
where $a_i^{(k+1)}=a_i^{(1)}$.
The perturbative and instanton parts, $Z_{[U(N)]^k}^{\rm (pert)}$ and $Z_{[U(N)]^k}^{\rm (inst)}$, are given by
\begin{eqnarray}
&&Z_{[U(N)]^k}^{\rm (pert)}=   \Bigl(\prod_{\alpha =1}^k \prod_{\substack{i,j=1 \\[+1pt] i\neq j}}^N Z_{\rm vec}^{(\rm pert)}(a_i^{(\alpha )} -a_j^{(\alpha )} )  
\Bigr)\Bigl(\prod_{\alpha =1}^k \prod_{\substack{i,j=1 \\[+1pt] i\neq j}}^N Z_{\rm mat}^{(\rm pert)} (a_i^{(\alpha )} -a_j^{(\alpha +1 )} ) \Bigr), \\
&& Z_{[U(N)]^k}^{\rm (inst)} 
= \sum_{Y^{(1)},\cdots ,Y^{(k)}} \exp\left(-\frac{8\pi^2}{g_d^2} \sum_{\alpha=1}^k |Y^{(\alpha )}|\right) \nonumber\\
&&\qquad\qquad\qquad \times 
 \prod_{\alpha =1}^k \prod_{i,j=1}^N 
 Z_{\rm vec}^{\rm (inst)} (a_i^{(\alpha )} -a_j^{(\alpha )} ;Y^{(\alpha )}_i ,Y^{(\alpha )}_j )
Z_{\rm mat}^{\rm (inst)} (a_i^{(\alpha )} -a_j^{(\alpha +1)} ;Y_i^{(\alpha )}, Y_j^{(\alpha +1)} ;\tilde{m}) .\nonumber
\end{eqnarray}

We can confirm the orbifold equivalence in the zero-instanton sector in the following way. 
At $g_p^2,g_d^2\ll 1$, one can use the saddle point method, as we have seen before. 
For the daughter theory, we introduce the spectral densities for the $k$ $U(N)$ gauge groups, $\rho^{(1)},\cdots ,\rho^{(k)}$. 
Then the saddle point equation of the parent and daughter theories coincide by taking the `democratic ansatz', 
\begin{eqnarray}
\rho^{(1)}(y) =\cdots = \rho^{(k)}(y) .
\end{eqnarray}
By substituting this ansatz, one obtains $F^{\rm (pert)}_p(\lambda,N) = kF^{\rm (pert)}_d(\lambda,N)$ for the zero-instanton sector 
even without using the detail of the spectral densities.

A generalization of the orbifold equivalence to the instanton sectors goes as follows, 
so far as the instanton number is of order one.  
We compare the partition functions of the parent and daughter theories
at each instanton sector.  
Once one fixes a sector to consider, 
at the leading order of the large-$N$ limit,
the eigenvalues describing instantons can be treated 
as probes and thus do not affect the distribution of other eigenvalues.
Then the free energies of the $U(kN)$ and $[U(N)]^k$ theories are respectively expressed as 
\eqref{eq:instanton_free_energy_U(kN)} and 
\begin{eqnarray}
 F^\text{(inst)}_{d, Y}
 &=&
 -N \sum_{\alpha=1}^k \sum_{b_i \in \text{(inst)}} 
 \int da\,\rho^{(\alpha)}(a) \log Z_{\rm vec}^\text{(inst)}(b_i^{(\alpha)},a,Y_i,\emptyset)
 \nonumber\\
 & &
 -
   N \sum_{\alpha=1}^k  \sum_{b_i \in \text{(inst)}} 
 \int da\,\rho^{(\alpha+1)}(a) \log Z_{\rm mat}^\text{(inst)}(b_i^{(\alpha)},a,Y_i,\emptyset; \tilde{m})
 +\sum_{\alpha=1}^k\sum_{b_i \in \text{(inst)}}\Delta_{\rm (pert)}^{(\alpha)}(b_i),  
 \nonumber\\
 \label{eq:instanton_free_energy_U(N)^k}
\end{eqnarray}
($\rho^{(k+1)}(a)=\rho^{(1)}(a)$) at the leading order in the large-$N$ limit.  
By substituting the democratic ansatz to \eqref{eq:instanton_free_energy_U(N)^k} 
and comparing it with \eqref{eq:instanton_free_energy_U(kN)} , we obtain $F_{p, Y}^{\rm (inst)} = kF_{d, Y}^{\rm (inst)}$ 
for each instanton sector.

Before closing this section, we remark on a subtle issue associated with the vacuum structure. 
As emphasized in \cite{Kovtun:2004bz}, the orbifold equivalence requires that
the vacuum structures of the parent 
and daughter theories be properly related. In the present case, because the numbers of instantons and anti-instantons 
are finite and of order one, it did not change the vacuum structure and the equivalence in the zero-instanton sector is naturally extended.  
When the number of instantons and anti-instantons is of order $N$, the vacuum structure 
in the large-$N$ limit is modified and hence 
careful identification of the right vacua is required. One has to assign the instantons and anti-instantons 
in the daughter theory `democratically' to the $k$ nodes, so that the instanton background becomes ${\mathbb Z}_k$ invariant 
and the democratic ansatz for the eigenvalues holds.

\section{Discussions}\label{sec:discussion}
Although we have used ${\cal N}=2^*$ gauge theory and its orbifold daughter theory 
preserving ${\cal N}=2$ supersymmetry for explicit demonstration, 
our calculation can be immediately generalized to other ${\cal N}=2$ theories. 
We note that we have considered ${\cal N}=2$ theories just because the free energies are calculable analytically.
As discussed in \cite{Azeyanagi:2012xj}, supersymmetry does not seem necessary. 

The very strongly coupled large-$N$ limit we have discussed in this letter 
may be useful for studying M-theory through gauge/gravity duality \cite{Maldacena:1997re}. 
Within the framework of string theory, gauge/gravity duality 
relates the classical gravity to the planar diagrams, and  
the $1/N$ expansion around the 't Hooft limit is identified with the string loop expansion. 
When it comes to M-theory, however, situation had not been clear because the 't Hooft coupling grows with $N$ 
where the dual gravity description turns to the eleven-dimensional  supergravity.
Our proposal for this issue is simple: the eleven-dimensional supergravity is also related to the planar sector 
in that the very strongly coupled large-$N$ limit simply picks it up at the leading order. 
As we have investigated here, 
it is true not only in the perturbative sector \cite{Azeyanagi:2012xj,Fujita:2012cf} but also in the instanton sectors. 
It is then expected that the total free energy like \eqref{total_free_energy} can be obtained by calculating 
the on-shell action on the gravity side.  We hope to report a development along this direction in near future. 

In the end of this letter, we speculate on a possible application of our result to large-$N$ QCD. 
In this theory, the beta function for the 't Hooft coupling becomes of order one in the 't Hooft limit, 
and hence the 't Hooft coupling at the UV cutoff should be taken $N$-independent. 
It however does not necessarily mean that instantons must obey a naive counting; 
the coupling constant in the instanton action should be evaluated at the characteristic energy scale of the instantons, 
which is the inverse of the radius of instantons. 
Therefore, as the inverse of the radius approaches the QCD scale, the 't Hooft coupling diverges and 
the very strongly coupled large-$N$ limit can be realized. 
It might then happen that small instantons are exponentially suppressed and the contribution 
with  $g^2=O(1)$ becomes dominant. 
It would be interesting if the results of the lattice simulations are reproduced in this way.

\section*{Acknowledgement}
We would like thank A.~Cherman, P.~H.~Damgaard, T.~Kimura, B.~Lucini and M.~\"{U}nsal for stimulating discussions and comments.
T.~A. is in part supported by JSPS Postdoctoral Fellowship for Research Abroad. 
He is also grateful to the Center for the Fundamental Laws of Nature at Harvard University for support. 
M.~Hanada is financially supported by the Hakubi Center for Advanced Research 
and JSPS Grant-in-Aid for Scientific Research (No.25800163).
He would like to thank the Center for Computational Science and Physics Department, Boston University,  
and the Aspen Center for Physics for hospitality during his stay.
M.~Honda is grateful to the Yukawa Institute for Theoretical Physics and the Hakubi Center for Advanced Research 
for hospitality and financial support.
Y.~M. and S.~S. are partially supported by Grant-in-Aid for JSPS Fellows (No.23-2195 and 23-7749).


\end{document}